\documentstyle[multicol,prb,aps]{revtex}

\begin{document}
\draft
\title{Complex impedance of a spin injecting junction}
\author{Emmanuel I. Rashba\cite{Rashba*}}    
\address{Department of Physics, The State University of New York at Buffalo, Buffalo, NY 14260}  
\date{December 21, 2001}
\maketitle
\begin{abstract} 
Theory of the ac spin injection from a ferromagnetic electrode into a normal conductor through a tunnel or Schottky contact is developed. Diffusion and relaxation of non-equilibrium spins results in a frequency-dependent complex impedance ${\cal Z}(\omega)$ controlled by the spin relaxation rates and the resistances involved. Explicit expression for ${\cal Z}(\omega)$ is presented. Experimental investigation of the frequency dependence of ${\cal Z}$ should allow measuring spin relaxation times in both conductors, their effective resistances, and also the parameters of the contact controling the spin injection. 
\end{abstract}
\begin{multicols}{2}
\narrowtext
Spin injection from magnetic contacts into semiconductor microstructures is one of the basic problems of the emerging field of spintronics.\cite{Wolf} Effective spin injection has been first achieved from semimagnetic semiconductors but only at low temperatures and in an external magnetic field. Reliability of the first reports\cite{HBYJ99,GSBLR99} on the observation of spin injection from ferromagnetic metals into semiconductors at the level of about 1\% has been questioned. The problem stems from the conductivity mismatch between a metal and a semiconductor, and the spin injection coefficient $\gamma$ at the level of only about $\gamma\sim \sigma_N/\sigma_F$ is expected from a ``perfect" Ohmic contact,\cite{Sch00} $\sigma_F$ and $\sigma_N$ being conductivities of a ferromagnetic aligner and a normal conductor, respectively. If the latter is a semiconductor and the former is a metal, then $\sigma_N \ll \sigma_F$ and, therefore, $\gamma \ll 1$. However, it has been proposed that employing a spin selective contact with the resistance $r_c\agt r_F, r_N$, where $r_F$ and $r_N$ are effective resistances of both conductors, can fix the problem.\cite{R00} With spin selective contacts of enlarged resistance, values of $\gamma\sim 10\%$ have been recently reported by different experimental groups.\cite{Plo01,HJ01,Jon01,Saf01} 

Therefore, a reliable and non-destructive control of the resistances $r_c, r_F$, and $r_N$ is an important problem. Physically $r_c$ is the resistance of a tunnel or a Schottky barrier that is strongly technology dependent. Effective resistances of both conductors depend not only on their bulk conductivities but also on the spin diffusion lengths in these materials, $L_F$ and $L_N$, as $r_F=L_F/\sigma_F$ and 
$r_N=L_N/\sigma_N$. Measuring the dc resistance of a junction provides only a single quantity, its total resistance $R$. In what follows, I find an explicit expression for the frequency dependence of the complex impedance 
${\cal Z}(\omega)$ of an F-N-junction in the diffusive regime and show that it can provide a lot of information on the parameters involved.

Let us consider a junction consisting of a ferromagnetic conductor (F) at $x<0$, a spin selective contact at $x=0$, and a normal conductor (N) at $x>0$. The theory is based on the system of standard equations\cite{R00,vS87,VF93,HZ97} relating the currents of up- and down-spin electrons $j_{\uparrow,\downarrow}(x,t)$ in the F region to their electrochemical potentials $\zeta_{\uparrow,\downarrow}(x,t)$ 
\begin{equation}
j_{\uparrow,\downarrow}(x,t)=
\sigma_{\uparrow,\downarrow}\nabla\zeta_{\uparrow,\downarrow}(x,t),
\label{eq1}
\end{equation}
these potentials being connected to the non-equilibrium concentrations $n_{\uparrow,\downarrow}(x,t)$ 
of both spins as
\begin{equation}
\zeta_{\uparrow,\downarrow}(x,t)=
(eD_{\uparrow,\downarrow}/\sigma_{\uparrow,\downarrow})
n_{\uparrow,\downarrow}(x,t)-\varphi_F(x,t).
\label{eq2}
\end{equation}
Here $\varphi_F(x,t)$ is the electric potential and $D_{\uparrow,\downarrow}$ are diffusion coefficients that are related to the temperature dependent densities-of-states $\rho_{\uparrow,\downarrow}=\partial n_{\uparrow,\downarrow}/\partial(e\zeta_{\uparrow,\downarrow})$ by Einstein equations $e^2D_{\uparrow,\downarrow}=
\sigma_{\uparrow,\downarrow}/\rho_{\uparrow,\downarrow}$. If the Poisson equation is substituted by the quasineutrality condition
\begin{equation}
n_{\uparrow}(x,t)+n_{\downarrow}(x,t)=0,
\label{eq3}
\end{equation}
then the continuity and charge conservation equations take a form
\begin{eqnarray}
\nabla j_{\uparrow}&(x,t)& =en_{\uparrow}(x,t)/\tau_s^F,\nonumber\\
J(t)&=&j_{\uparrow}(x,t)+j_{\downarrow}(x,t),
\label{eq4}
\end{eqnarray}
where $J(t)$ is the total current and $\tau_s^F$ is the spin relaxation time.

In symmetric variables 
\begin{eqnarray}
\zeta_F(x,t)&=&\zeta_\uparrow(x,t)-\zeta_\downarrow(x,t), \nonumber\\
Z_F(x,t)&=&{\case 1/2}[\zeta_\uparrow(x,t)+\zeta_\downarrow(x,t)],\nonumber\\
j_F(x,t)&=&j_\uparrow(x,t)-j_\downarrow(x,t),
\label{eq5}
\end{eqnarray}
the basic equations are
\begin{eqnarray}
\nabla^2\zeta_F(x,t)&=&\zeta_F(x,t)/L_F^2+\partial_t\zeta(x,t)/D_F,\nonumber\\
\nabla Z_F(x,t)&=&
-(\Delta\sigma/2\sigma_F)\nabla \zeta_F(x,t)+J(t)/\sigma_F,
\label{eq6}
\end{eqnarray}
\begin{equation}
j_F(x,t)=2(\sigma_\uparrow\sigma_\downarrow/\sigma_F)\nabla\zeta_F(x,t)
+(\Delta\sigma/\sigma_F)J(t),
\label{eq7}
\end{equation}
where $D_F=(\sigma_\downarrow D_\uparrow +\sigma_\uparrow D_\downarrow)/\sigma_F$ is the ``bispin" diffusion coefficient, $\sigma_F=\sigma_\uparrow+\sigma_\downarrow$,
$\Delta\sigma=\sigma_\uparrow-\sigma_\downarrow$,
 and $L_F^2=D_F\tau_s^F$. Similar equations operate in the N region.

Neglecting the spin relaxation in the contact, the boundary conditions at the contact are 
\begin{eqnarray}
j_F(0,t)&=&j_N(0,t)\equiv j(0,t),\nonumber\\
j_{\uparrow,\downarrow}(0,t)
&=&\Sigma_{\uparrow,\downarrow}(\zeta^N_{\uparrow,\downarrow}(0,t)
-\zeta^F_{\uparrow,\downarrow}(0,t)),
\label{eq8}
\end{eqnarray}
where $\Sigma_{\uparrow,\downarrow}$ are spin selective conductivities of the contact, $\zeta^F_{\uparrow,\downarrow}(0,t)$ and $\zeta^N_{\uparrow,\downarrow}(0,t)$ are the values of $\zeta^{F(N)}_{\uparrow,\downarrow}(x,t)$ at both sides of the contact, and similarly for $j_{\uparrow,\downarrow}(0,t)$. The first of equations (\ref{eq8}) ensures the continuity of the spin-polarized current, and the second relates the partial spin polarized currents to the discontinuities in the corresponding electrochemical potentials. 

If the time dependence of the ac signal is chosen as $\exp(-i\omega t)$, than in the diffusion equation  (\ref{eq6}) for the F and N regions the spin diffusion lengths should be changed to
\begin{equation}
{1\over{L^2_F(\omega)}}={1\over{L^2_F}}(1-i\omega\tau_s^F),~~
{1\over{L^2_N(\omega)}}={1\over{L^2_N}}(1-i\omega\tau_s^N),
\label{eq9}
\end{equation}
i.e., they become complex and frequency dependent. Nevertheless, the procedure of solving the equations for $\zeta_F(x,t)$ and $\zeta_N(x,t)$ with boundary conditions of Eq.~(\ref{eq8}) is essentially the same as in the dc regime. Because $\zeta_F(x\rightarrow -\infty,t)\rightarrow 0$ and $\zeta_N(x\rightarrow\infty,t)\rightarrow 0$, the difference $Z_N(x_N,t)-Z_F(x_F,t)$ tends to the potential drop $\varphi(x_F,t)-\varphi(x_N,t)$ between these points for large $x_F$ and $x_N$. However, it is important to emphasize that the conditions under which Eqs.~(\ref{eq6}) and (\ref{eq7}) are valid in the dc and the ac regimes are very different. In the dc regime they can be derived without using the quasineutrality condition of Eq.~(\ref{eq3}), while in the ac regime Eq.~(\ref{eq3}) should be explicitly employed.

An explicit equation for the dc resistance $R$ of a F-N-junction that is most convenient for our purposes is as follows: 
\begin{eqnarray}
 R=&R_{\rm eq}&+R_{\rm n-eq},~~R_{\rm eq}=\Sigma^{-1}=(\Sigma_\uparrow+\Sigma_\downarrow)^{-1},\nonumber \\
 R_{\rm n-eq}&=&{1\over{r_{FN}}}\{ r_N\left[r_c(\Delta\Sigma/\Sigma)^2
+r_F(\Delta\sigma/\sigma_F)^2 \right]\nonumber \\
&+&r_c r_F\left[(\Delta\Sigma/\Sigma)-(\Delta\sigma/\sigma_F)\right]^2 \},
\nonumber\\
r_{FN}&=&r_F+r_N+r_c. 
\label{eq10}
\end{eqnarray}
It consists of two parts. The equilibrium resistance $R_{\rm eq}$ is the limit of $R$ when $L_N, L_F\rightarrow 0$, i.e., when spin non-equilibrium can be neglected both in the F and N regions. It depends only on the contact resistance. 
$R_{\rm n-eq}$ is the non-equilibrium resistance controlled by spin diffusion and relaxation. The parameters of the spin selective contact entering the non-equilibrium resistance $R_{\rm n-eq}$ are
\begin{equation}
\Delta\Sigma=\Sigma_\uparrow-\Sigma_\downarrow,~
r_c=\Sigma/4\Sigma_\uparrow\Sigma_\downarrow.
\label{eq11} 
\end{equation}
It is worth mentioning that two different effective resistances of the contact, $R_{\rm eq}$ and $r_c$, appear in the equilibrium and non-equilibrium parts of the total resistance $R$, respectively.

It is an important property of $R_{\rm n-eq}$ that it is always positive, 
$R=R_{\rm n-eq}>0$, as it has been already concluded in Ref.\onlinecite{R00}. Therefore, there is a considerable difference between the spin injection and bipolar injection through a $p$-$n-$junction. Bipolar injection always contributes to the conductivity, and a large resistance of the deplicion region plays no essential role.\cite{Sh49} For a F-N-junction, the resistance $R_{\rm n-eq}$ resulting from spin injection is added to $R_{\rm eq}$. Large positive magnetoresistance of a spin-injecting device has been recently observed in weak magnetic fields and attributed to the magnetic alignment of semimagnetic electrodes enhancing spin injection.\cite{Sch01}

In the ac regime, the impedance ${\cal Z}(\omega)$ can be found from 
Eq.~(\ref{eq10}) by the transformation 
$L_F, L_N \rightarrow L_F(\omega), L_N(\omega)$ of equation (\ref{eq9}):
\begin{eqnarray}
 {\cal Z}_{\rm n-eq}(\omega)&&\nonumber\\
={1\over{r_{FN}(\omega)}}
&\{& r_N(\omega)\left[r_c(\Delta\Sigma/\Sigma)^2
+r_F(\omega)(\Delta\sigma/\sigma_F)^2 \right]\nonumber \\
&+&r_c r_F(\omega)\left[(\Delta\Sigma/\Sigma)-(\Delta\sigma/\sigma_F)\right]^2 \},
\label{eqZ}
\end{eqnarray}
where
\begin{equation}
r_F(\omega)=L_F(\omega)/\sigma_F,~~r_N(\omega)=L_N(\omega)/\sigma_N .
\label{eqr}
\end{equation}
Eq.~(\ref{eqZ}) is the basic result of the paper.

The imaginary part of ${\cal Z}(\omega)$ results in a reactive conductivity that can be compared to the diffusion capacitance of a $p$-$n-$junction.\cite{Sh49}  The frequency dependence of ${\rm Re}\{{\cal Z}\}$ and ${\rm Im}\{{\cal Z}\}$ provides a useful tool for measuring different parameters of a F-N-junction. Because $r_c$ does not depend on $\omega$, frequency dependence comes exclusively through $r_F(\omega)$ and $r_N(\omega)$. Some general regularities follow directly from equations (\ref{eq9}) and (\ref{eqZ}). Spin relaxation times $\tau_s^F$ and $\tau_s^N$ in a ferromagnetic aligner and a semiconductor microstructure, respectively, may differ by several orders of magnitude. Usually $\tau_s^F\ll \tau_s^N$. Therefore, two different scales, $(\tau_s^F)^{-1}$ and $(\tau_s^N)^{-1}$, should emerge in the frequency dependence of ${\cal Z}(\omega)$. This property opens an opportunity for measuring these parameters separately, and through them to evaluate $r_F$ and $r_N$. At high frequencies, $\omega\gg (\tau_s^F)^{-1},(\tau_s^N)^{-1}$,  the diffusion contribution to $\cal Z$ vanishes and ${\cal Z}\rightarrow R_{\rm eq}$.

It is instructive to consider the low- and high-frequency regimes in more detail. Expansion of ${\rm Im}{\cal Z}(\omega)$ in $\omega$ starts with a linear in $\omega$ term, and the positive sign of ${\rm Im}\{{\cal Z}\}$ suggests that ${\rm Im}\{{\cal Z}^{-1}\}$ can be considered as the conductivity $\omega C_{\rm diff}$ of a capacitor connected in parallel to the resistor $R$:
\begin{eqnarray}
C_{\rm diff}&=&\biggl\{\tau_s^N r_N\left(r_c{{\Delta\Sigma}\over\Sigma}
+r_F{{\Delta\sigma}\over{\sigma_F}}\right)^2 \nonumber\\
&+&\tau_s^Fr_F\left[r_c{{\Delta\Sigma}\over\Sigma}
-(r_c+r_N){{\Delta\sigma}\over{\sigma_F}}
\right]^2
\biggr\}/2R^2r_{FN}^2.
\label{eq12}
\end{eqnarray}
It is seen from Eq.~(\ref{eq12}) that $C_{\rm diff}>0$ for all parameter values. Capacitance arising from the diffusion of non-equilibrium carriers is typical of injection devices and is well known for $p$-$n-$junctions.\cite{Sh49} However, the dependence of the low-frequency diffusion capacitance on the relaxation time $\tau$ is rather different for these systems. The square root dependence, $C_{\rm diff}\propto \tau^{1/2}$, is typical of $p$-$n-$junctions. Diffusion capacitance $C_{\rm diff}$ of a F-N-junction follows this law only when $r_c\ll r_N, r_F$, i.e., when spin injection into a semiconductor is strongly suppressed. In the opposite limit $r_c\gg r_N, r_F$ that is of major interest for spin injecting devices, it follows from Eq.~(\ref{eq12}) that $C_{\rm diff}\propto\tau_s^{3/2}$. Depending on the relative magnitudes of $r_F$ and $r_N$, different combinations of $\tau_s^F$ and $\tau_s^N$ can appear in $C_{\rm diff}$. Therefore, large $\tau_s^N$ typical of semiconductor microstructures should enlarge $C_{\rm diff}$ considerably.\cite{KA98}

In the high frequency regime, when $\omega\tau_s^F,~ \omega\tau_s^N\gg 1$, the resistances $r_F(\omega)$ and $r_N(\omega)$ are small and one can expand ${\cal Z}_{\rm diff}(\omega)$ in
$r_F(\omega)/r_c,~ r_N(\omega)/r_c\ll 1$. Because in this limit 
$L_F(\omega)\approx (1+i)L_F/{\sqrt{2\omega\tau_s^F}}$ and similarly for
$L_N(\omega)$, the real and imaginary parts of ${\cal Z}_{\rm diff}(\omega)$ are nearly equal and
\begin{eqnarray}
R_{\rm n-eq}(\omega)&=&{1\over{\sqrt{2\omega}}}\nonumber\\
&\times& 
\left[{{r_N}\over{\sqrt{\tau_N}}}\left({{\Delta\Sigma}\over\Sigma}\right)^2
+{{r_F}\over{\sqrt{\tau_F}}}
\left({{\Delta\Sigma}\over\Sigma}-{{\Delta\sigma}\over{\sigma_F}}\right)^2\right]
\label{eq13}
\end{eqnarray}
\begin{equation}
C_{\rm diff}(\omega)=R_{\rm n-eq}(\omega)/\omega R_{\rm eq}^2.
\label{eq14}
\end{equation}
It is seen from Eqs.~(\ref{eq13}) and (\ref{eq14}) that both $R_{\rm n-eq}(\omega)$ and $C_{\rm diff}(\omega)$ remain positive even in the high frequency limit.

Side by side with the diffusion capacitance $C_{\rm diff}$, there always exists a geometrical capacitance $C_{\rm geom}\approx \varepsilon/4\pi X$, where $X$ is the contact thickness and $\varepsilon$ is the dielectric permeability. $C_{\rm geom}$ depends on the specific geometry of a tunnel or Schottky contact and cannot be found in a general form. However, it does not depend on the frequency $\omega$ and 
this property is critical for eliminating the by-passing effect of the geometric capacitance $C_{\rm geom}$ and measuring ${\cal Z}_{\rm diff}(\omega)$. It is the frequency dependence of the diffusion impedance ${\cal Z}_{\rm diff}(\omega)$ with two characteristic frequencies $(\tau_s^F)^{-1}$ and $(\tau_s^N)^{-1}$ that should facilitate separating the diffusion and geometric contributions to the total impedance ${\cal Z}(\omega)$. Because Eq.~(\ref{eqZ}) provides explicit expression for ${\cal Z}_{\rm diff}(\omega)$, a quantitative comparison of experimental data with the theory is possible in the wide region of frequencies.

In a similar way, a frequency-dependent complex spin injection coefficient 
$\gamma (\omega)=j_F(\omega)/J(\omega)$ can be found:
\begin{equation}
\gamma(\omega)=[r_F(\omega)(\Delta\sigma/\sigma_F)
+r_c(\Delta\Sigma/\Sigma)]/r_{FN}(\omega).
\label{eq15}
\end{equation} 
It describes the frequency dependence of the magnitude of the spin-polarized current $j_F(t)$ and the phase shift between $j_F(t)$ and $J(t)$. Non-equilibrium spins related to the current $j_F(t)$ can also be detected in optical experiments. Optical data should not be obscured by the by-pass current flowing through
the geometric capacity $C_{\rm geom}$. 

Because the frequency dependence of the impedance ${\cal Z}(\omega)$ comes completely from the frequency dependence of the diffusion lengths $L_F(\omega)$ and $\L_N(\omega)$ that are the parameters of the bulk, a similar approach can be applied to the impedance of  F-N-F-junctions. However, the equation for the dc resistance $R$ of a F-N-F-junction is much more cumbersome than Eq.~(\ref{eq10}). 

In conclusion, an explicit expression for the complex impedance 
${\cal Z}(\omega)$ of a spin injecting junction is presented. The diffusion contribution to ${\cal Z}(\omega)$ shows strong frequency dependence with two characteristic frequencies corresponding to the inverse spin relaxation times in the ferromagnetic emitter and the normal conductor into which spins are injected. Electrical and (or) optical detection of the time dependent spin injection should allow measuring spin relaxation times and effective resistances controlling the efficiency of spin injection. 

Support from DARPA/SPINS by the Office of Naval Research Grant N000140010819 is gratefully acknowledged.

\end{multicols}
\end{document}